\documentclass[12pt, a4paper]{article}

\begin{document}
\title{Exploiting Randomness in \\
Quantum Information Processing}
\author{Chiu Fan Lee\thanks{c.lee1@physics.ox.ac.uk} 
\ and \ Neil F. Johnson\thanks{n.johnson@physics.ox.ac.uk}
\\
\\ Center for Quantum Computation and Physics Department \\ Clarendon
Laboratory, Oxford University \\ Parks Road, Oxford OX1 3PU, U.K.}

\maketitle

\abstract{
We consider how randomness can be made to play a useful role in quantum
information processing - in particular, for decoherence control and
the implementation of quantum algorithms. For a two-level system in
which the decoherence channel is non-dissipative, we show that 
decoherence suppression is possible if memory is present in the
channel. Random switching between two potentially harmful
noise sources can then provide a source of stochastic control. Such random
switching can also be used in an advantageous way for the implementation
of quantum algorithms.}

\newpage

\section{Introduction}
Randomness and noise are typically seen as having a detrimental
effect on the coherent evolution of a
quantum system, and hence on the ability of the system to process quantum
information \cite{Chuang}. 
For classical systems, the advent of phenomena such as
stochastic resonance, Brownian ratchets and the Parrondo effect, have
shown that noise may indeed play a helping role after all \cite{PHA00}. 
This opens up the intriguing
question of whether randomness can play a useful role in quantum
systems, in particular given the widespread current interest in quantum
information schemes such as quantum computation \cite{Chuang}. 

Here we investigate how the intrinsic randomness of 
an open quantum system might actually be used to our advantage for
quantum information processing. First we
consider the role of randomness in suppressing and controlling decoherence
(Secs. 2 and 3). For a
two-level system in which the decoherence channel is non-dissipative, we
show that  suppresion of decoherence is possible if memory is present in
the channel. Random switching between two potentially harmful noise
sources, can then provide a source of stochastic control. 
Second, we show how random switching can be used in an advantageous way
for the implementation of quantum algorithms (Sec. 4).

\section{Stochastic decoherence}
Decoherence is a unique quantum phenomenon which results in a decay 
of the off-diagonal elements in a density matrix. Simply put, it 
is the following process:
\[
\rho_0 :=\left( \begin{array}{cc}
a & b \\ b^*  & c
\end{array} \right)
\stackrel{t}{\mapsto}
\left( \begin{array}{cc}
a & b' \\ b'^*  & c
\end{array} \right)
\]
where $|b'| < |b|$. 
This results in the decay of superpositions of states into a probablistic
mixture.

\subsection{Non-dissipative system}
We now study decoherence in a two-level system
under the physical assumptions that
the channel is (i) non-dissipative, and (ii) isolated, i.e. no
entanglement is allowed between the system and the environment.
This decoherence can be thought of as arising from the observers' limited
knowledge of the channel, e.g. due to uncontrollable classical
fluctuations. If we further assume discrete time evolution for
simplicity, the above assumptions imply that the final density  matrix
can be written as
$
\rho_n= U_n \cdots U_1 \rho_0 U_1^\dag \cdots U_n^\dag
$
where
\[
U_k = 
\left( \begin{array}{cc}
e^{-i \theta_k /2} & 0 \\ 0  & e^{i \theta_k /2}
\end{array} \right).
\]
Decoherence can only occur if our knowledge
of $\theta_k$'s is uncertain. To prove that such a lack of knowledge will almost surely lead
to decoherence, let us assume that 
the phase kicks $\theta$ are independent and identically distributed
with probability distribution 
$P(\theta)$. We will then have
\[
\rho_n =
\left( \begin{array}{cc}
a & b\gamma^n e^{- i n\phi} \\ b^\ast \gamma^n e^{i n\phi} & c
\end{array} \right)
\]
where 
\[
\gamma e^{\pm i\phi}:= \int_{\theta_{lo}}^{\theta_{hi}} e^{\pm i \theta}
 P(\theta) d \theta. 
\]
Hence, $|\gamma| \leq 1$ and the equality is satisfied if and only if $P(\theta) = \sum_k p_k
\delta(\theta - a - k)$ for some constant
$a$
\cite{Luk70}. This condition will only be met in exceptional circumstances - therefore
stochastic decoherence will essentially always arise in such a system. 

We now let $\tau_0$ be the interaction time and set 
$n=t/ \tau_0$.
Letting $P(\theta)$ be $\frac{exp(\theta / \omega \tau_1)}{\omega \tau_1}$ for $\theta\geq
0$,  we find $\gamma = \sqrt{1+ \omega^2 \tau_1^2}$
and $\phi = \arctan (\omega \tau_1)$. This coincides with the main result in Ref. \cite{BOT00}
where the constant
$\tau_1$ corresponds to the
``time width of each event''
\cite{BOT00}. On the other hand, by setting
$P(\theta) := \frac{1}{\sigma \sqrt{2 \pi}} \exp \Big[
-\frac{(\theta-\mu)^2}{2\sigma^2} \Big]$ where 
$\mu:=\sin (\omega/ \lambda)$ and $\sigma^2:= 2(1-\cos (\omega /\lambda))$,
we recover the result in Ref. \cite{Mil91} if we identify
$\lambda$ with
the ``fundamental time of the universe'' \cite{Mil91}.
This observation makes sense because our assumptions
are the most general ones. Indeed for any decay factor of the form
$(\gamma e^{i \phi})^{t/\tau_0}$ with $\gamma <1$, we could pick 
$P(\theta) :=  \frac{1}{\sigma \sqrt{2 \pi}} \exp \Big[
-\frac{(\theta-\mu)^2}{2\sigma^2} \Big]$ with 
$\mu:= \phi$ and  $\sigma^2:= -2 \ln \gamma$.

Given that stochastic decoherence will almost always be present in a real-world quantum
system, in accordance with the discussion above, one can ask whether a method can be devised
to control it. Unfortunately, we can show that there
is actually no way to suppress the stochastic 
decoherence discussed above {\em if} the channel has no memory.
The justification of this statement is as follows. Let us assume that such an operation is
feasible and let us call it ${\cal F}$, where
\[
{\cal F}: 
\left( \begin{array}{cc}
a & b  \\ 
b^\ast & c \end{array} \right)
\mapsto
\left( \begin{array}{cc}
a & b'  \\ 
b'^\ast & c \end{array} \right)
\]
and where $|b'| > |b|$ for a non-empty set of $b$ values. 
We require that ${\cal F}$ be non-dissipative,
i.e. we don't allow dissipation in exchange for decoherence suppression.
Hence
\[
{\cal F}
\left(\left( \begin{array}{cc}
1 & 0  \\ 
0 & 0 \end{array} \right) \right) =
\left( \begin{array}{cc}
1 & 0  \\ 
0 & 0 \end{array} \right) \ \ ; \ \
{\cal F}
\left( \left( \begin{array}{cc}
0 & 0  \\ 
0 & 1 \end{array} \right) \right) =
\left( \begin{array}{cc}
0 & 0  \\ 
0 & 1 \end{array} \right).
\]
Letting
\[
{\cal F}
\left( \left( \begin{array}{cc}
0 & 1  \\ 
0 & 0 \end{array} \right) \right)=
\left( \begin{array}{cc}
\alpha & \beta  \\ 
\gamma & \delta \end{array} \right) \ \ ; \ \
{\cal F}
\left( \left( \begin{array}{cc}
0 & 0  \\ 
1 & 0 \end{array} \right) \right)=
\left( \begin{array}{cc}
\alpha' & \beta'  \\ 
\gamma' & \delta' \end{array} \right),
\]
we see that for some $b$, $|\beta +e^{i \theta} \beta'| >1$ where
$e^{i \theta} = b^* /b$. 
But if we now take 
$\rho_0 := 
\frac{1}{\sqrt{2}}\left( \begin{array}{cc}
1 & e^{i \theta/2}  \\ 
e^{-i \theta/2} & 1 \end{array} \right)$,
then 
\[
{\cal F}(\rho_0) = 
\frac{1}{\sqrt{2}}\left( \begin{array}{cc}
1   & \beta +e^{i \theta} \beta' \\
(\beta +e^{i \theta} \beta')^*
 & 1 \end{array} \right)
\]
This latter quantity ${\cal F}(\rho_0) $ will not have positive eigenvalues because the
off-diagonal elements have norms greater than 1, 
thereby contradicting the fact that ${\cal F}$
is a superoperator.
In other words, the above demonstration shows
that there is no `coherence booster' -
even for particular states.
 
\subsection{Dissipative system}
Through the following example, we illustrate how a dissipative 
system may be modelled using a similar stochastic process. Consider the following channel:
\[
{\cal E}(\rho) = pE_0 \rho E_0^\dag + pE_1 \rho E_1^\dag + (1-p)R_z \rho
R_z^\dag
\]
where
\[
E_0:=
\left( \begin{array}{cc}
1&0 \\ 0& \sqrt{1-\alpha}
\end{array} \right) \  ,  \
E_1:=
\left( \begin{array}{cc}
0& \sqrt{\alpha} \\ 0&0
\end{array} \right) \  ,  \
R_z:=
\left( \begin{array}{cc}
e^{-i \theta/2} & 0 \\ 0& e^{i \theta/2}
\end{array} \right)
\]
Hence
\[
{\cal E}(\rho) =
\left( \begin{array}{cc}
1-(1-\alpha p)(1-a) & b [p\sqrt{1-\alpha} + (1-p)e^{-i \theta}] \\ 
b^\ast [p\sqrt{1-\alpha} + (1-p)e^{i \theta}] & c (1-\alpha p)
\end{array} \right).
\]
Keeping $p$ fixed and assuming $\alpha, \theta$ to be Gaussian and independent,
gives
\[
\int \int
{\cal E}(\rho) P_{ad}(\alpha) P_{pd}(\theta) d \alpha d \theta =
\]
\[
\left( \begin{array}{cc}
1-(1-p\sqrt{\frac{4 \lambda_{ad}}{\pi}})(1-|a|^2) &
ab^\ast [p(1-\sqrt{\frac{\lambda_{ad}}{\pi}})+(1-p)e^{-\lambda_{pd}}] \\
a^\ast b [p(1-\sqrt{\frac{\lambda_{ad}}{\pi}})+(1-p)e^{-\lambda_{pd}}] &
|b|^2 (1-p\sqrt{\frac{4 \lambda_{ad}}{\pi}})
\end{array} \right)
\]
where we have assumed $\lambda_{ad} \ll 1$, and a Taylor expansion
has been used on the diagonal terms before integrating.
In particular, if it is known that  the relevant timescales satisfy $T_1 \geq T_2/2$ (see e.g.
Ref.
\cite{AL87}),  then we need
\[
p \leq\frac{1-e^{-\lambda_{pd}}}{1-e^{-\lambda_{pd}}+\sqrt{\lambda_{ad} / \pi}}
\ .
\]

\section{Control of decoherence via randomness}
Decoherence control is crucial to the success of quantum computation. We
now study the possibility of
controlling stochastic decoherence using further randomness.

\subsection{A vector-rotating game}
We start by introducing a classical vector-rotating game which shows a Parrondo-like
effect \cite{PHA00}. This particular game motivates much of the later
development of decoherence control and the discussion of algorithms.

\noindent {\em Game A:\ } Consider a wheel with a vector 
drawn from the center to the circumference, i.e. the vector is a radial
line.  Suppose the vector is originally vertical (i.e. $\theta=0$) and the
player plays by calling a robot (A) to rotate the  wheel. The robot can only
rotate the wheel by
$0$, $2\pi/3$ or
$4\pi/3$ radians, with equal probabilities. The player wins if the vector
ends up in the  upper-half of the circle (i.e. $-\pi/2\leq \theta\leq
\pi/2$) and he loses otherwise. The game is continued by rotating the wheel
from the previous position, i.e. without restoring the vector to the
vertical position. The stationary states are such that the vector will
end up at $\theta=0$, $2\pi/3$ or
$4\pi/3$
with equal probabilities. Therefore
this game is losing for the player and the rate of losing is $1/3$. In
Parrondo's original game, the losing rate is smaller (i.e. $-2\epsilon$ where
$\epsilon\ll 1$).

\noindent {\em Game B:\ } This is the same as game
A, except that the robot (B) can now only rotate the wheel by 
$0$, $2\pi/7$,
$4\pi/7$, $6\pi/7$, $8\pi/7$, $10\pi/7$, $12\pi/7$,  with
equal probabilities. Similar analysis as that for game A shows the
player's losing rate is $1/7$. In
Parrondo's original game, the losing rate is again smaller 
(i.e. $-11\epsilon/5$ where 
$\epsilon\ll 1$).

\noindent {\em Game $A\oplus B$:\ } The player now plays a combined game
in which he randomly selects either A or B at each timestep. 
Operationally, one of the robots A or B is
selected at random to rotate the wheel at each timestep. Simple geometric
analysis shows that the vector can now end up in
$3\times 7=21$ different orientations, $11$ of which are
winning. 
The corresponding  $21\times21$ transition matrix is doubly-stochastic and
so the stationary distribution will be equally distributed among these $21$
positions. Therefore the player now wins with probability
$11/21\approx 0.5238 >1/2$. In
Parrondo's original game, the winning rate was $1/80 -
21\epsilon/10$ as compared to the present, larger rate of $1/21$. 
It turns out there is nothing special
about the numbers $3$ and $7$ chosen for this implementation. The games A
and B are originally losing simply because $3=7=3 \bmod 4$, and the combined
game becomes winning because
$3\times 7 =1 \bmod 4$. Therefore, the above vector-rotating implementation
of Parrondo's effect works equally well for all $m,n$ such that $(m,n)=1$ and
$m=n=3\bmod 4$. By the same method, we can therefore construct two losing
games with rates
$-1/m<0$ and
$-1/n<0$ such that when they are combined at random, we obtain a
winning game with rate
$1/mn>0$. One could also extend the Parrondo scheme to include random
combinations of  {\em any even} number of games.

\subsection{Stochastic control}
We showed earlier that it is impossible to control decoherence
if the noise does not have any ``memory''.
This then leads us to consider correlated phase kicks. Depending on 
the particular model employed, some of the control methods devised elsewhere
might also work \cite{Closedloop,  Openloop}. 
However, we choose here to focus on a 
stochastic suppression of decoherence which mimics some form of Parrondo effect
\cite{PHA00}.

Motivated by the classical vector-rotation game we introduced earlier,
we consider two probability distributions $P_A, P_B$ which are
correlated to the previous rotated angle ($\theta_1$) in the following manner:
\[
P_A(\theta_2|\theta_1) =
\left\{ \begin{array}{ll}
\frac{1}{3}[ \delta(0) +\delta(-\pi/2) + \delta(\pi/2 )]
 & , \ \ \ \ \theta_1 \in \{-\pi/2, 0, \pi/2 \} \\
\delta(0) & ,\ \ \ \ {\rm otherwise} 
\end{array} \right.
\]
\[
P_B(\theta_2|\theta_1) =
\left\{ \begin{array}{ll}
\frac{1}{3}[ \delta(\epsilon)
 +\delta(-3\pi/4) + \delta(\pi/4 )]
 & , \ \ \ \ \theta_1 \in \{-3\pi/4, \epsilon, \pi/4\} \\
\delta(\epsilon) & ,\ \ \ \ {\rm otherwise}. 
\end{array} \right.
\]
If $P_A$ is the only noise in the system and if we assume the initial
angle of rotation is $0$, we will have
\[
P_A(\theta_n, \ldots, \theta_1)= \prod_i P_A(\theta_i)
\]
as the $\theta_i$'s always lie in the set  $\{-\pi/2, 0,\pi/2\}$.
Therefore,
\[
\gamma_A e^{\pm i\phi_A}:= \int e^{\pm i \theta}
 P_A(\theta) d \theta 
= \frac{1}{3}.
\]
Similarly,
\[
\gamma_B e^{\pm i\phi_B}:= \int e^{\pm i \theta}
 P_B(\theta) d \theta 
= \frac{1}{3}e^{i \epsilon} 
\]
with $\gamma_A=\gamma_B=1/3$.

Combining the two probability distributions at random gives
\[
P(\theta_2|\theta_1) =
\left\{ \begin{array}{ll}
\frac{1}{2}\delta(\epsilon)
 + \frac{1}{6} [\delta(0) +\delta(-\pi/2) + \delta(\pi/2 )] 
 & , \ \theta_1 \in \{-\pi/2, 0,\pi/2 \} \\
\frac{1}{2}\delta(0)
 + \frac{1}{6} [\delta(\epsilon)
 +\delta(-3\pi/4) + \delta(\pi/4)] 
 & , \ {\rm otherwise}. \\
\end{array} \right.
\]
Since $R_z(\theta)R_z(\phi)=R_z(\phi)R_z(\theta)$, we can write $\rho_n$ as
\[
\int R_z(\theta_1) \cdots \int R_z(\theta_n) \rho_0
R_z^\dag (\theta_n) P(\theta_n|\theta_{n-1}) d \theta_n
\cdots R_z^\dag (\theta_1) P(\theta_1)
d \theta_1 
\]
We now define the following functions
recursively:
\begin{eqnarray*}
f_1(\theta) &:=& \int e^{i \phi}P(\phi| \theta) d \phi \\
f_{k+1}(\theta) &:=&
\int e^{i\phi}f_k(\phi) P(\phi| \theta) d \phi
\end{eqnarray*}
for $1\leq k\leq n$. Here $\rho = f_n(0)$, assuming that the initial angle is 0.

For the combined probability distribution $P$ above, we see that
the angles of rotation can only take on six values, 
$\{ -3\pi/4 , -\pi/2, 0,  
 \epsilon, \pi/4, \pi/2 \}$. Furthermore, we can
calculate the $f_k$'s to be the following:
\[
f_1:
\left\{ \begin{array}{ll}
\{-\pi/2, 0, \pi/2 \} & \mapsto e^{i \epsilon}/2 + 1/6 \\
\{ -3\pi/4 , \epsilon, \pi/4 \}
& \mapsto 1/2 + e^{i \epsilon}/6
\end{array} \right.
\]
\[
f_{k+1}:
\left\{ \begin{array}{ll}
\{-\pi/2, 0, \pi/2 \} & \mapsto 
\frac{1}{2} f_k(0) + \frac{1}{6} f_k( \epsilon) \\
\{  -3\pi/4 ,\epsilon,  \pi/4 \}
& \mapsto \frac{1}{2} f_k(\epsilon) + \frac{1}{6} f_k(0).
\end{array} \right.
\]
Letting $\epsilon$ go to zero and writing $e^{i \epsilon}$ as
$1 + {\cal O}(\epsilon)$, we see that the $f_k$'s always output 
$2/3 + {\cal O}(\epsilon)$. An immediate consequence is that
$\rho^{(10)}_n$ has an exponential decay factor of $2/3 +
{\cal O}(\epsilon)$. This is an {\em improvement} over the value $1/3$, which is the decay 
factor if  
we were to consider noise $P_A$ and noise $P_B$ separately. This result is
reminiscent of the Parrondo effect discussed earlier for classical systems \cite{PHA00}. 
We note in passing that Mancini {\it et al} ~\cite{MVT01} have also devised a
stochastic scheme to control quantum coherence. 
Although these authors invoke a memoryless modulation of the cavity length, their model
is dissipative. More specifically, their only source of decoherence is the loss of photons -
hence their results are fundamentally different from
the present case of a {\em non}-dissipative channel.

\section{A stochastic algorithm}
We now turn to an example involving quantum algorithms, in which cooperating with
randomness may be a better strategy than trying to fight it. We consider 
a game where the player's goal is to obtain (i.e. measure with a high
probability) a fixed, unknown number
$\alpha$ in as few timesteps as possible. Here $0
\leq
\alpha \leq 2^n-1$. The initial
state has the form
$|\psi\rangle=\sum^{2^n-1}_{x=0}
\frac{1} {\sqrt{2^n}}|x\rangle$.  In this game, an infinite sequence of
operators
$
\hat{O}_1
\cdots \hat{O}_m \cdots$ will be applied to $|\psi \rangle$. The player
decides when to stop the sequence, i.e. he has the freedom
to choose
$m$ such that $|\psi_f\rangle = \hat{O}_m \cdots \hat{O}_1|\psi\rangle$. The
payoff is then determined by a measurement in the computational basis of
$|\psi_f\rangle$. The game is winning if the player possesses a strategy
that wins with probability $>1/2$,  and is losing
otherwise. This game incorporates strategic  moves, since the
set of strategies used by the player to decide the duration of the game 
are equivalent to the set of natural numbers
${\bf N}$.

\noindent {\em Game A:\ } Here $\hat{O_i} = \hat{A}$ for all $i$, 
 where $\hat{A}(|x \rangle ) = (-1)^{\delta_{x \alpha}} |x \rangle$. 
Geometrically, $\hat{A}$ reflects the vector $|\psi\rangle$ about $|\alpha
\rangle$. Since $\hat{A}^2=I$, the player's freedom in choosing when to
stop the game will always reduce to just one of the following two
scenarios: 
$|\psi_f \rangle = \hat{A} |\psi \rangle$ or
$|\psi_f \rangle = |\psi \rangle$. Unfortunately for the player,
the payoff $|\langle \alpha | \hat{A}|\psi \rangle|^2 = |\langle \alpha |\psi
\rangle|^2 = \frac{1}{2^n}$ which is less than $1/2$ for $n
\geq2$. Therefore the player does not possess a winning strategy,
hence game A is losing for him. 

\noindent {\em Game B:\ }  Here $\hat{O_i}=\hat{B}$ for all $i$, where
$\hat{B}:=2|\psi \rangle \langle \psi | - I$. 
Geometrically, $\hat{B}$ reflects $|\psi \rangle$ about itself.
Again, the player has the freedom to decide how many $\hat{B}$ are applied
to the input state before measurement. However since $\hat{B}|\psi \rangle
=|\psi
\rangle$, the player can have no influence in determining the payoff
in this game. The  game is hence losing for him because the payoff $|\langle
\alpha |
\psi
\rangle|^2 = \frac{1}{2^n}$ which is less than $1/2$.

\noindent {\em Game $A\oplus B$:\ } The player combines games A and B at
random. By this we mean $\hat{O}_i = \hat{A}$ or $\hat{B}$ with equal
probability. Once
again, the player has the freedom to decide when to stop the sequence and
hence do the measurement.
Since $\hat{A}^2=\hat{B}^2 =I$ and
$\hat{B}|\psi \rangle=|\psi\rangle$, any given finite
sequence
$\hat{O}_i$ will always produce a final state with the following form: 
$|\psi_f\rangle =(\hat{B})\hat{A}\hat{B}\cdots \hat{A}\hat{B}\hat{A}|\psi
\rangle$. Now, numerical calculation suggests that for $m=4k$,
\begin{eqnarray*} |\psi_f\rangle &=& \hat{O}_{m} \cdots
\hat{O}_1|\psi\rangle \\ &=&
\overbrace{(\hat{B}\hat{A})\cdots (\hat{B}\hat{A})}^{k}|\psi\rangle.
\end{eqnarray*} 
It can also be seen that $\hat{B}
\circ
\hat{A} =
\hat{G}
$ where $\hat{G}$ is Grover's operator \cite{Grover,Chuang}. Hence
a winning strategy for the player is to choose to stop
after the $4k$-th operation where $k=\lceil \pi \sqrt{2^n}/4 \rceil$.
The winning probability is $>1/2$, and hence
we see that this combined game is winning for the
player.

\section{Conclusion}
We have discussed potentially useful roles for randomness in quantum information
processing - in particular, decoherence control and quantum algorithms.
The counter-intuitive conclusion is that such randomness/noise
might be of direct use in the quantum regime, as opposed to being a guaranteed nuisance. We
hope that the present work serves to simulate further research in this fascinating area.

\vskip0.5in CFL thanks NSERC (Canada), ORS (U.K.) and
Clarendon Fund (Oxford) for financial support. NFJ thanks L.
Quiroga and F. Rodriguez for discussions.

\end{document}